# Satyendra Nath Bose: Quantum statistics to Bose-Einstein condensation

**Golam Ali Sekh[1*], Benoy Talukdar [2]**
[1] Department of Physics, Kazi Nazrul University, Asansol 713340, WB, India
[2] Department of Physics, Visva-Bharati University, Santiniketan 731235, WB, India

*E-mail: golamali.sekh@knu.ac.in

### Abstract

Satyendra Nath (S.N.) Bose is one of the great Indian scientists. His remarkable work on the black body radiation or derivation of Planck's law led to quantum statistics, in particular, the statistics of photon. Albert Einstein applied Bose's idea to a gas made of atoms and predicted a new state of matter now called Bose-Einstein condensate. It took 70 years to observe the predicted condensation phenomenon in the laboratory. With a brief introduction to the formative period of Professor Bose, this research survey begins with the founding works on quantum statistics and, subsequently, provides a brief account of the series of events terminating in the experimental realization of Bose-Einstein condensation. We also provide two simple examples to visualize the role of synthetic spin-orbit coupling in a quasi-one-dimensional condensate with attractive atom-atom interaction.

## I. INTRODUCTION

At the end of nineteenth century, properties of physical systems would be studied by using (i) classical mechanics, (ii) Maxwell's theory of electromagnetism and (iii) thermodynamics. The developments in (i) - (iii) made people believe that ultimate description of nature had been completed. However, at the turn of the twentieth century such a belief was challenged on two major fronts. First, Albert Einstein developed the special theory of relativity in 1905 and general theory of relativity in 1915. Both these revolutionary theories had profound impact on classical mechanics. In the special theory, the Newtonian formulation of mechanics was shown to be an approximation that applies only at low velocities. The Newtonian concept of an absolute frame of reference as well as the assumption of the separation of space and time was shown to be invalid at high velocities. The general theory of relativity superseded Newton's law of gravitation by providing a geometrical theory for the origin of gravitational force [1].

The other profound developments that led to revolutionary impacts on classical mechanics were quantum physics and quantum field theory formulated by Bohr, Sommerfeld, de Broglie, Heisenberg, Born, Schrödinger and Dirac. This conceptual revolution in physics took place during the first three decades of the twentieth century. The main objective was to explain several microscopic phenomena such as black-body radiation, photoelectric effect, atomic stability and atomic spectroscopy. Classical concepts were inadequate to provide their correct description; we needed a new theory - the so-called quantum theory. The origin of this theory is perhaps embedded in a talk given by Max Planck on December 14, 1900 to the German Physical society on the continuous spectrum of the frequencies of light emitted by an ideal heated body or the so-called black body [2]. Here Planck's immediate concern was to provide a radiation formula that can account for experimentally confirmed prediction of the black-body spectrum. He considered the black body as an ensemble of charged oscillators and derived a formula that reduces to Wien and Rayleigh-Jeans radiation laws in appropriate limits. As is well known, the Rayleigh-Jeans radiation formula when integrated over all frequencies leads to ultraviolet catastrophe. Planck's law gives a way out from this crisis. Since the radiation law as given by Max Planck is based on an educated guess, it was felt that Planck's formula should be derived from the first principle of statistical mechanics. In this context an unknown Indian-Satyendranath Bose, a young physicist from Dacca University (now in Bangladesh) provided a derivation of the Planck's law without reference to classical electrodynamics. The judgement of history on Bose's paper is that it not only laid the foundation stone of quantum statistics but also justified the photon concept of light that Einstein had championed since 1905. Einstein developed Bose's concept further, extended it to monatomic ideal gases, and predicted what is known as the Bose-Einstein condensation. As named by Paul Dirac, particles obeying Bose statistics has come to known as bosons. According to Abraham Pais 'The paper by Bose is the fourth and last of the revolutionary papers of the old quantum theory, the other three being by Planck, Einstein and Bohr [3]. Significantly enough, Bose's name is one of the six which under graduate physics students come across in the course of statistical mechanics - the others being Maxwell, Boltzmann, Einstein, Fermi and Dirac [4].

Satyendra Nath Bose was not an institution builder like Meghnad Saha, Homi Janngi Bhaba, Prasanta Mahalanobis or Shanti Swarup Bhatnagar. Throughout his life he was a professor with profound interest in different branches of science, statistics and mathematics, literature and music. He was very generous, gentle and, particularly, not caring about the glamorous aspects of science. It is, therefore, an interesting curiosity to look back into the formative period of Professor Bose and envisage a pedagogic study to visualize how Bose statistics led to our current understanding of the so-called Bose-Einstein condensate which has effectively changed our current understanding of matter [5].

## II. REMEMBERING THE EARLY LIFE OF BOSE

Satyendranath Bose was born in Calcutta (now called Kolkata) on the first January, 1894 in a high caste Hindu family with two generations of English education behind him. Both his grandfather, Ambika Charan, and father, Surendra Nath, were Government employees in British India. Satyendranath had an inborn talent and would have flourished under any circumstances. But it was a lucky coincidence that he found a congenial atmosphere.

Schooling of little Satyen began at the age of five. His family was then living at north Calcutta. First, he was admitted to 'Normal School' close to their residence and then shifted to the famous Hindu School which had a glorious tradition behind it. Although Satyendranath had varied interest, he was particularly strong in mathematics. The mathematics teacher of the school, Upendranath Bakshi, was a legend. He was quick to recognize the signs of genius in the boy. Once, in a test examination, he gave Satyen 110 marks out of 100; his argument was that, in the answer script, Satyen did not skip any of the alternatives. Mr. Bakshi even believed that one day Satyen would become a great mathematician like Laplace or Cauchy [6].

In the entrance examination of 1909 Satyen stood fifth in order of merit. In addition to mathematics, he did very well in Sanskrit, History and Geography. But he opted for the science course and joined the intermediate science class at the Presidency College. In the intermediate examination of 1911, Styendranath stood first and his illustrious colleagues Meghnad Saha (coming from Dacca College) and Nikhil Ranjan Sen secured the second and third positions respectively. All of them joined the B. Sc. class in the Presidency College and opted for mixed Mathematics. In the B. Sc. Examination of 1913 Satyendranath Bose stood first, Meghnad Saha second, and Nikhil Ranjan third, all in the first class. The same result was repeated in the M. Sc. mixed mathematics examination of 1915 except that Nikhil Ranjan did not appear in the examination in that year. The bright Satyendranath was now ready for a career.

## III. BEGINNING THE CARRIER

After completing the formal education in schools and colleges it was quite natural for Satyen to look for the prospect open before him. In those days jobs were difficult to get. But situation began to change as Sir Asutosh Mookherjee, the mathematician Vice Chancellor of Calcutta University introduced post graduate teaching program at the University. He immediately needed a band of teachers to run the program. In 1916 both Satyendranath and Meghnad were appointed as lecturers in the Applied Mathematics Department. But neither of them felt comfortable with the then Ghosh Professor of the Department, Dr. Ganesh Prasad. With kind permission of Sir Asutosh, both of them were transferred to the Physics Department although their formal training in Physics was up to B.Sc. level only. Bose and Saha were entirely self-taught in Physics. They studied modern Physics on their own and, remarkably enough, they translated Einstein's papers on the theory of relativity from German to English [7]. During 1920s the situation in the Physics Department of the Science College at Calcutta was becoming rather uncomfortable due to inadequacy of technical resources. Meanwhile, Bose was looking for a better opportunity. At that time a new university was established at Dacca and the authorities there were looking for competent teachers. Bose was offered readership. When Sir Asutosh came to know this, he expressed his willingness to increase Bose's salary. But Bose had already given his words to accept the appointment at Dacca. In 1921 he joined Dacca University. Mr. P. J. Hartog, the vice chancellor of the University gave Bose the task of building a new Department – including setting up of laboratories and teaching advance courses in Physics for B. Sc. Honours and M. Sc. students. Meanwhile, the library was also being equipped with books and journals. Bose taught thermodynamics and Maxwell's theory of Electromagnetism. Just as his first group of students graduated in 1923, Bose received a letter from the University authority notifying that his appointment will not be extended beyond a year. The reason behind such a decision was a conflict between government of India and the provincial government of Bengal regarding fund allocation for the University. This led Bose to a awkward position to keep his appointment. It was under this troubled situation that he wrote his famous paper on the derivation of Planck's law. The story regarding publication of the paper is well documented in the scientific literature. He sent the article for publication to Philosophical Magazine in the beginning of 1924. After six months the editors informed him that the referees had given negative reports. He sent the rejected paper to Albert Einstein. Einstein was impressed; he himself translates it from English to German and submitted for publication in Zeitschrift für Physik with an added note, "In my opinion Bose's derivation of the Planck's formula signifies an important advance. The method used also yields the quantum theory of ideal gas, as I will work out in detail elsewhere." It is perhaps the second sentence which contains the germ of Bose-Einstein condensation observed in the laboratory [8 – 10] after seventy years since the publication of Bose's paper [11]. In this context it will not be an exaggeration to say that it required the genius of Einstein to realise the far-reaching consequence of the work by Bose. During a friendly visit to Dacca in March 1924, Saha brought to Bose's attention about the new attempts by Wolfgang Pauli, and by Albert Einstein and Paul Ehrenfest to derive Planck's law. Saha's visit, perhaps, provided further impetus to Bose for thinking about the interaction of radiation with matter. In fact, this led Bose writing a second paper that he again sent to Einstein. We shall now present Bose's derivation of Plank's law. Needless to say, we shall begin by considering historically significant discoveries that played a key role to explain the spectrum of black body radiation as visualized by Max Planck.

## IV. BLACK-BODY RADIATION, PLANCK'S LAW AND BOSE STATISTICS

### (a) The observed Black-body spectrum

It is a common experience that all material bodies when heated emit radiation. The spectrum of black-body radiation represents one of the early experimental results, the theoretical explanation of which ultimately led to quantum ideas. Experimentally, the black-body radiation spectrum was first studied by Tyndall [12]. There are two important terms that are commonly used to characterize the nature of the radiating substance. These are the so-called emissive and absorptive powers. A black body is made up of a substance whose absorptive power is unity. The term black body was coined by Kirchhoff [13].

Figure 1 gives the schematic diagram of the original Tyndall experiment. It consists of the black-body light source, a collimating slit and lens, a prism and focusing lens, and light sensor mounted on a rotating arm. A rotary motion sensor measures the angle. The incandescent light source that emits light through a small cavity is a perfect emitter. When light from the black body is cast through a prism, the observed spectrum is continuous. Different wavelengths of light will project to different angles.

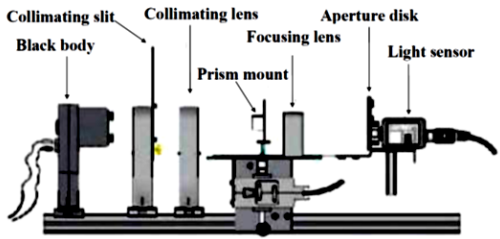

**FIGURE 1:** Diagram for the setup of Tyndall experiment

In this experiment, parallel light rays travel through the collimating lens, which allows the light rays to remain parallel. Passing through the prism, the light rays refract and project in front of the aperture slit over the light sensor. The light sensor detects and records the light intensity as voltage. So, by measuring the voltage as a function of angle, one can find the intensity of radiation in the spectrum as a function of wavelength.

Figure 2 displays a typical black-body spectrum giving the intensity as a function of wave length. The curves in this figure clearly show that the black-body spectrum is temperature dependent. The intensity of radiation at any given temperature tends to zero at both shorter and longer wavelengths and has a maximum in between. The maximum tends towards shorter wavelength as temperature increases.

### (b) Attempts for theoretical explanation

Calculating the black-body curve was a major challenge in theoretical physics during the late nineteenth century because of the following.

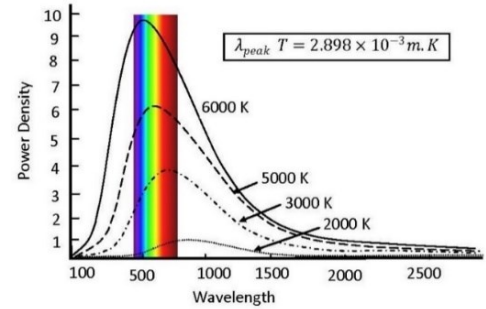

**FIGURE 2:** Black-body spectrum. Power density is measured in the unit of $10^3\ Watts/m^3$ and wavelength in $nm$

(1) The radiation spectrum is not influenced by factors like the substance of emitting body or condition of its surface. The black-body spectrum is then a pure and ideal case. If one could describe the energy distribution of this ideal case then one would learn something about radiation process in all cases.

(2) It is the basic thermodynamic state of light in which radiation is in thermal equilibrium with a given temperature. Light is an electromagnetic field. The black-body radiation shows that the continuous electromagnetic field can have temperature dependence. This point was not physically realizable at that time.

In view of the above there were attempts to explain the nature of the black-body spectrum by using thermodynamical methods. The thermodynamical consideration of perfect gas led to far-reaching consequences like the discovery of temperature radiation. In this context we note the following similarities between a perfect gas and black-body radiation both of which may be supposed to be confined in enclosures.

(i) In the kinetic theory we assume a perfect gas as being an assembly of particles having all velocities from zero to infinity and moving in all directions.

(ii) In the case of black-body radiation, the radiation also proceeds in all directions and is composed of waves of all lengths.

(iii) The gas molecules in a perfect gas exert pressure on the wall.

(iv) In the case of radiation also, the light waves carry momenta and exert pressure when they are incident on the walls.

Thus, it was tempting to find analogy between black-body radiation and a perfect gas. Studies in the distribution of energy in the black-body spectrum were begun by Wien [14]. The radiation emitted by a black body is not confined to a single wavelength but spreads over a continuous spectrum. The problem was to determine how the energy is distributed over different wavelengths. Wien showed that $E_\lambda d\lambda$ i. e. the amount of energy contained in the spectral region included within the wavelengths $\lambda$ and $\lambda + d\lambda$ emitted by a black body at a temperature $T$ is of the form

$$E_\lambda d\lambda = \frac{A}{\lambda^5} f(\lambda T) d\lambda. \qquad (1)$$

Using $\lambda = c/v$, Eq. (1) can be written in the equivalent form

$$v_v dv = Bv^3 \varphi\left(\frac{v}{T}\right) dv \qquad (2)$$

with $v_v$, the energy density of the radiation having frequency $v$. This expression was obtained from purely thermodynamical consideration applied on a Gedanken experiment which involves a spherical enclosure having perfectly reflecting walls capable of slowly moving outwards. The enclosure was assumed to be maintained at some temperature and a small black-body of negligible heat capacity was placed inside it. Formula in Eq. (1) was derived by considering thermal equilibrium between the two. As a consequence of adiabatic expansion Wien could deduce

$$\lambda T = \text{Constant.} \qquad (3)$$

Equation (3) is often called the displacement law [2,14]. The physical interpretation of the law is that if radiation of a particular wavelength at a certain temperature is adiabatically altered to another wavelength, then the temperature changes in the inverse ratio. Wien also made assumptions regarding the mechanism for emission and absorption of radiation. The radiation inside a hallow enclosure was supposed to be produced by a resonator of molecular dimension and the frequency of the wave emitted is proportional to the kinetic energy of the resonators. The resonators were supposed to obey the Boltzmann statistics [15]. This consideration converts the energy distribution law in Eq. (1) in the form

$$E_\lambda d\lambda = \frac{A}{\lambda^5} e^{-c_2/\lambda T} d\lambda, \qquad (4)$$

where $c_2 = \alpha \, c/k_B$ and $\alpha$, constant.

Let us now try to see to what extent the formula in Eq. (3) can explain the experimental black-body spectrum. For given values of $c_2$ and $A$, $E_\lambda$ vanishes at $\lambda = 0$ and $\lambda = \infty$. Thus, it appears that the energy distribution law in (3) is a good candidate to explain black-body spectra as given in figure 2. More specifically,

(i) Paschen [16] working with light of short wavelength verified that Wien's formula fits the data for short waves.

(ii) On the other hand, Lummer and Pringshiem [17] working with long waves and high temperature found considerable disagreement of the theoretical values from the experimental results.

In the above context we note that thermodynamical consideration could not give any improved expression for $f(\lambda T)$ to fit the experimental data. Wien discovered his energy distribution law in 1893. After seven years Lord Rayleigh [18] attempted to find an energy distribution function by the use of classical electromagnetic theory. The work was completed by Sir James Jeans [19]. The law discovered by them goes by the name Rayleigh-Jeans law. To derive the law, they considered a black body chamber in the form of a parallelepiped with perfectly reflecting walls. Also, they assumed that there is a black particle inside. In the course of time the enclosure will be filled with stationary waves of all lengths, for the particle emits radiation which is reflected back by the wall. The reflected and incident waves interfere and form stationary waves.

The black body chamber is filled with diffuse radiation of all frequencies between 0 to $\infty$. Using the above picture Lord Rayleigh found out the number of possible wave motion having their frequencies between $v$ and $v + dv$. The number of vibrations per unit volume was calculated as $\frac{4\pi v^2}{c^3} dv$.

Since electromagnetic waves are transverse, they can be polarized and each polarized component is independent of the other. Thus, the required number of vibrations per unit volume is $\frac{8\pi v^2}{c^3} dv$. Converting frequency into wavelength the number becomes

$$\frac{8\pi}{\lambda^4} d\lambda. \qquad (5)$$

The energy of each vibration is $k_B T$. In view of this the energy distribution law obtained by Rayleigh and Jeans is given by

$$E_\lambda d\lambda = \frac{8\pi}{\lambda^4} k_B T \, d\lambda. \qquad (6)$$

The distribution in Eq. (6) can account for the long wavelength part of the black-body spectrum. For smaller values of $\lambda$, $E_\lambda$ tends to $\infty$. This is the so-called ultraviolet catastrophe. This implies that if the black body chamber is initially filled with infrared radiation, finally it will be filled up with ultraviolet radiation. Thus, we see that neither of the radiation formulas, one given by Wien and other given by Rayleigh and Jeans, can explain the black-body spectrum. Therefore, explanation of the black-body spectrum using theoretical consideration assumed the status of an unsolved problem.

### (c) Planck's Law of black-body radiation

Planck [20] imagined that a black-body radiation chamber is filled up not only with radiation, but also with the molecules of a perfect gas. At that time, the exact mechanism of generation of light by atomic vibrations or of absorption of light by atoms and molecules was unknown. Planck, therefore, introduced resonators of molecular dimensions as the via media between radiation and gas molecules. These resonators absorb energy from the radiation, and transfer energy partly or wholly to the molecules when they collide with them. In this way thermodynamical equilibrium is established.

The resonators introduced by Planck were dipole oscillators which may be described as Hertzian oscillators of molecular dimensions such that the density $v_v$ of radiation of frequency $v$ could be written as

$$v_v = \frac{8\pi v^2}{c^3} E_v , \qquad (7)$$

where $v_v$ is the mean energy of a resonator emitting the radiation. According to classical idea $E_v = k_B T$. As a result, the expression in Eq. (7) gives the Rayleigh-Jeans law which is inconsistent with the experimental data. Planck abandoned the hypothesis of continuous emission of radiation by resonators, and assumed that they emit energy only when the energy is an integral multiple of certain minimum energy $\epsilon$. As we know currently, this assumption is equivalent to light quantum hypothesis of Einstein [21]. In any case, let us try to calculate the mean energy of these resonators. The probability that a resonator will possess the energy $E$ is $exp(-E/k_B T)$. Let $N_0$, $N_1$, $N_2$, ..., $N_r$, .... be the number of resonators having energies $0, \epsilon, 2\epsilon, 3\epsilon, ...., r\epsilon, ....$ Then we have

$$N = N_0 + N_1 + N_2 + \cdots + N_r + \cdots, \qquad (8)$$

and

$$E = \epsilon[N_1 + 2N_2 + 3N_3 + \cdots + rN_r + \cdots] \qquad (9)$$

with

$$N_r = N_0 e^{-r\epsilon/k_B T} \qquad (10)$$

Using Eq. (10) in (8) we get

$$N = \frac{N_0}{1 - \exp\left[-\frac{\epsilon}{k_B T}\right]}. \qquad (11)$$

Again using Eq. (10) in (9) we get,

$$E = \epsilon N_0 \frac{\exp\left[-\epsilon/k_B T\right]}{(1 - \exp\left[-\epsilon/k_B T\right])^2}. \qquad (12)$$

Dividing Eq. (12) by (11) we can write

$$\frac{E}{N_0} = \frac{\epsilon}{\exp\left[-\frac{\epsilon}{k_B T}\right] - 1}. \qquad (13)$$

We know from the law of equipartition of energy that the mean energy of a resonator is $k_B T$. This result agrees with that given in Eq. (13) only at an extremely high temperature. The energy assigned to the resonators, namely, $0, \epsilon, 2\epsilon, 3\epsilon, \ldots, r\epsilon, \ldots$ correspond to the light quantum hypothesis in that resonators can have only discrete set of energies. Thus, the mean energy given in Eq. (13) is a quantum law. Using Eq. (13) in Eq. (7) we get the energy density inside the enclosure as

$$u_\nu d\nu = \frac{8\pi\nu^2}{c^3} \frac{\epsilon \, d\nu}{\exp\left[-\frac{\epsilon}{k_B T}\right] - 1}. \qquad (14)$$

Comparing Eq. (14) with (2), the Wien's distribution law, we see that $\epsilon$ must be proportional to $\nu$. In view of this, Planck took $\epsilon = h\nu$, where $h$ is the so-called Planck's constant. Thus

$$u_\nu d\nu = \frac{8\pi h\nu^3}{c^3} \frac{d\nu}{\exp\left[-\frac{h\nu}{k_B T}\right] - 1}. \qquad (15)$$

Using $d\nu = -\frac{c}{\lambda^2} d\lambda$, we get

$$u_\lambda d\lambda = \frac{8\pi hc}{\lambda^5} \frac{d\lambda}{\exp\left[-\frac{ch}{\lambda k_B T}\right] - 1}. \qquad (16)$$

Equation (16) is known as the Planck's law of radiation. In the short wavelength limit Eq. (16) gives the Wiens distribution law and in the long wavelength limit we get Rayleigh-Jeans law. Thus, the Planck's formula could explain the black-body spectrum satisfactorily.

Planck's hypothesis that resonators can have only discrete energies resolved the essential mysteries of the black-body radiation. The subsequent works of Einstein on the photoelectric effect and of Compton on the scattering of X-rays established the discrete or quantum nature of radiation [22]. Thus, Planck's work is a statement of quantum hypothesis of light. A quantum of radiation goes by the name photon. It was then natural to look for derivation of Planck radiation formula by treating the black-body radiation as a gas of photons in a similar way as Maxwell derived his distribution law for a gas of conventional molecules. But a gas of photons differs radically from a gas of conventional molecules because Maxwell's molecules are classical objects while photon is a purely quantum mechanical concept and is thus indistinguishable. In this context Bose derived a Statistics for indistinguishable particles (quantum statistics) and made use of it to deduce Planck's formula. In his historic paper of 1924, Bose [11] treated black-body radiation as a gas of photons; however, instead of considering the allocation of the "individual" photons to the various energy states of the system, he fixed his attention on the number of states that contained a particular number of photons. We shall try to elucidate this point in some detail.

**(d) Bose Statistics**
Let us try to calculate the distinct number of ways in which $N_s$ indistinguishable particles can be distributed in $A_s$ indistinguishable boxes. We refer to these boxes as cells since they represent minimum volume in the phase space when we apply this method for the derivation of Planck's law. Let the cells be designated as $x_1, x_2, x_3 \ldots \ldots, x_{A_s}$. A particular distribution can be represented by

$$x_1^\alpha x_2^\beta \ldots \ldots \ldots x_{A_s}^\gamma, \qquad (17)$$

where $\alpha, \beta, \ldots \ldots \ldots, \gamma$ are the number of particles in the cells $x_1, x_2, \ldots \ldots x_{A_s}$ respectively. Clearly,

$$\alpha + \beta + \cdots + \gamma + \cdots = N_s. \qquad (18)$$

Now consider the product

$$(x_1^0 + x_1^1 + x_1^2 + \cdots x_1^r + \cdots)(x_2^0 + x_2^1 + x_2^2 + \cdots x_2^r + \cdots) \ldots \ldots \ldots (x_{A_s}^0 + x_{A_s}^1 + x_{A_s}^2 + \cdots x_{A_s}^r + \cdots), \qquad (19)$$

where each factor consists of an infinite number of terms. In this product we have all possible combinations of the powers of $x_1, x_2, x_3 \ldots \ldots, x_{A_s}$. Hence the number of ways of distributing $N_s$ particles in the $A_s$ cells is equal to the number of those terms of type (17) for which the condition (18) is satisfied.

Now let $x_1 = x_2 = x_3 = \cdots = x_{A_s} = x$. The number of combinations in which the $N_s$ indistinguishable particles can be distributed in $A_s$ cells is equal to the coefficient of $x^{N_s}$ in this expression

$$(x^1 + x^2 + x^3 + \cdots + x^r + \cdots)^{A_s} = (1 - x)^{-A_s}. \qquad (20)$$

Thus, the number of ways in which $N_s$ number of indistinguishable particles can be distributed in $A_s$ indistinguishable cells is

$$\frac{(A_s + N_s - 1)!}{(A_s - 1)! \, N!} \qquad (21)$$

This is the so-called Bose statistics.

**(e) Bose's deduction of Planck's law**
A black-body chamber may be supposed to be full of photons in thermal equilibrium. The problem of finding spectral distribution of energy then reduces to that of finding the number of photons possessing energy $h\nu$ in a black-body chamber having temperature $T$. Bose realized the problem in this way and gave a very powerful method for the derivation of Planck's law [23]. According to quantum hypothesis, a radiation of frequency $\nu$ consists of photons of energy $h\nu$. The photons move in all possible directions with the constant velocity $c$ and momentum $h\nu/c$. Thus

$$p_x = \frac{h\nu_x}{c}, p_y = \frac{h\nu_y}{c} \text{ and } p_z = \frac{h\nu_z}{c} \qquad (22)$$

such that

$$p_x^2 + p_y^2 + p_z^2 = \frac{h^2\nu^2}{c^2}. \qquad (23)$$

Let us now find out the phase space volume described by the photons within the energy layers $h\nu_s$ and $h(\nu_s + d\nu_s)$. This is given by

$$G_s = \int \ldots \int dx \, dy \, dz \, dp_x dp_y dp_z \qquad (24)$$

with $G_s = V \frac{4\pi h^3 \nu_s^2}{c^3} d\nu_s$. This is the phase space volume at the disposal of the photons in the energy range $h\nu_s$ and $h(\nu_s + d\nu_s)$. But each photon has a phase volume $h^3$. Thus, the number of cells per unit volume

$$A_s d\nu_s = \frac{4\pi \nu_s^2 d\nu_s}{c^3}. \tag{25}$$

Since two photons are distinguished by their state of polarization from each other, instead of (20) we must write

$$A_s d\nu_s = \frac{8\pi \nu_s^2 d\nu_s}{c^3}. \tag{26}$$

The result in Eq.(26) is in agreement with that in (7) obtained by Rayleigh. Let the number of photons of frequency between $\nu_s$ and $(\nu_s + d\nu_s)$ be denoted by $N_s d\nu_s$. We have then to find out the number of ways in which the $N_s d\nu_s$ oscillators can be distributed amongst the $A_s d\nu_s$ cells. We make supposition that each cell may contain 1,2,3, ..., $r$, ..... upto $N_s d\nu_s$ photons. Then we get

$$W = \prod_s \frac{(A_s + N_s) d\nu_s!}{A_s d\nu_s! \, N_s d\nu_s!}, \tag{27}$$

according to Bose statistics in Eq. (21), as the probability of $N_s d\nu_s$ indistinguishable particles to be distributed in $A_s d\nu_s$ cells. Using $W$ in the Boltzmann relation between entropy and probability

$$S = k_B \ln W \tag{28}$$

we get

$$S = k_B \sum \ln \frac{(A_s + N_s) d\nu_s!}{A_s d\nu_s! \, N_s d\nu_s!}. \tag{29}$$

To obtain the law of distribution, Bose optimized the entropy subject to the constraint

$$E = \sum_s (N_s d\nu_s) h\nu_s, \text{ Constant} \tag{30}$$

which imply that the total energy $E$ of the photon gas is conserved. From (28) we have

$$\delta \sum_s [(A_s + N_s) \ln(A_s + N_s) - A_s \ln A_s - N_s \ln N_s] = 0, \tag{31}$$

using Stirling's formula

$$\ln n! \approx n \ln n - n. \tag{32}$$

Accommodating the energy constraint

$$\sum \nu_s \delta N_s = 0 \tag{33}$$

through the method of Lagrange undetermined multiplier, we get

$$N_s = \frac{A_s}{e^{\alpha \nu_s} - 1}. \tag{34}$$

Here $\alpha$, is the undetermined multiplier. Using $\alpha = \frac{h}{k_B T}$, Eq. (34) becomes

$$N_s = \frac{A_s}{e^{h\nu_s/k_B T} - 1}. \tag{35}$$

From Eq. (26) and the fact that the energy density

$$\rho_{\nu_s} d\nu_s = N_s h\nu_s d\nu_s, \tag{36}$$

we get the Planck's result

$$\rho d\nu = \frac{8\pi \nu^2 d\nu}{c^3} \frac{h\nu}{e^{h\nu/k_B T} - 1}. \tag{37}$$

We remember that Planck deduced this law by making use of hypothetical molecular resonators of discrete energies. On other hand, the treatment of Bose explicitly demonstrates that the concept of photons can be used to derive the Planck's law. In this way Bose's treatment provided a definitive proof for the light quantum hypothesis.

## V. ON THE REALIZATION OF BOSE-EINSTEIN CONDENSATION

### (a) Einstein's quantum theory of ideal gas

We have seen how Bose derived his statistics for the probability of distributing $N_s$ indistinguishable particles in $A_s$ cells. This remarkable result provided a natural basis to deduce Planck's radiation law for the explanation of black-body spectrum without taking recourse to the use of classical electromagnetic theory. Bose's derivation of Planck's formula is an application of his statistics to massless particles. Einstein [24] recognized that the method employed by Bose can also be generalized to deal with massive particles and thus have a quantum theory for the ideal gas. To derive this quantum mechanical theory let us consider a gaseous system of $N$ noninteracting indistinguishable particles confined in a volume $V$ and sharing a given energy $E$. The statistical quantity of interest in this case is the number of distinct microstates $\Omega(N, V, E)$ accessible to the system characterized by $(N, V, E)$. For large $V$, the single particle energy levels in the system are very close to one another. Thus, we may divide the energy spectrum into a large number of groups of levels which may be referred to as the energy cells. This is schematically shown in figure 3.

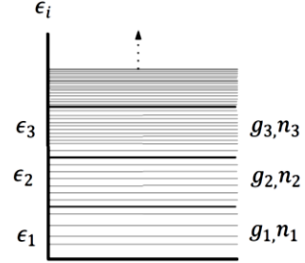

**Figure 3:** Distribution of particles in the energy cells

Let $n_i$ be the number of particles in the $i^{th}$ cell. Clearly, the set $\{n_i\}$ will satisfy the conditions

$$\sum_i n_i = N \tag{38}$$

and

$$\sum_i n_i \, \varepsilon_i = E \tag{39}$$

with $\varepsilon_i$, the average energy of a level. If $W\{n_i\}$ stands for the number of distinct microstates associated with the distribution set $\{n_i\}$, then the number of distinct microstates accessible to the system will be given

$$\Omega(N, V, E) = \sum_{\{n_i\}} W\{n_i\}. \tag{40}$$

The summation in Eq. (40) is taken over all distinct sets that obey the conditions in Eqs. (38) and (39). Again if $w(i)$ is the number of distinct microstates associated with the $i^{th}$ cell of the spectrum, then

$$W\{n_i\} = \prod_i w(i). \tag{41}$$

Bose statistics tells us that

$$w(i) = \frac{(n_i + g_i - 1)!}{n_i! (g_i - 1)!}, \tag{42}$$

where $g_i$ is the number of levels in the $i^{th}$ cell. Then

$$W\{n_i\} = \prod_i \frac{(n_i + g_i - 1)!}{n_i! (g_i - 1)!}. \tag{43}$$

The entropy of the system is given by

$$S(N, V, E) = k \ln \Omega(N, V, E) = k \ln \left[ \sum_{\{n_i\}} W\{n_i\} \right]. \tag{44}$$

The expression in Eq. (44) can be replaced by

$$S(N, V, E) \approx k \ln W \{n_i^*\}, \tag{45}$$

where $\{n_i^*\}$ is the distribution set that maximizes the number $W\{n_i\}$; the numbers $n_i^*$ are clearly the most probable values of the distribution number $n_i$. The maximization should be carried out under the constraints implied by Eqs. (38) and (39). Thus

$$\delta \ln W \{n_i\} - [\alpha \sum_i \delta n_i + \beta \sum_i \varepsilon_i \delta n_i] = 0, \tag{46}$$

where $\alpha$ and $\beta$ are Lagrange's undetermined multipliers.

Now

$$ln\, W\,\{n_i\} = ln \prod_i w(i) = \sum_i ln\, w\,(i) \qquad (47)$$

Using Stirling's formula $ln\, x\,! = x\, ln\, x - x$, we can write from Eqs. (42) and (47)

$$ln\, W\,\{n_i\} \approx \sum_i \left[ n_i\, ln\left(\frac{g_i}{n_i}+1\right) + g_i\, ln\left(1+\frac{n_i}{g_i}\right)\right], (48)$$

From Eqs. (46) and (48) we get

$$n_i^* = \frac{g_i}{e^{\alpha+\beta\varepsilon_i}-1}. \qquad (49)$$

Equivalently,

$$\frac{n_i^*}{g_i} = \frac{1}{e^{\alpha+\beta\varepsilon_i}-1}. \qquad (50)$$

may be interpreted as the most probable number of particles per energy level in the $i$ th cell. It is important to note that the final result in Eq. (50) is totally independent of the manner in which the energy levels of the particles are grouped into the cells so long as the number of levels in each cell is sufficiently large. From Eqs. (45), (48) and (49), the entropy of the gas is given by

$$\frac{S}{k} = \sum_i \left[ n_i^*(\alpha+\beta\varepsilon_i) - g_i\, ln\left(1-e^{-\alpha-\beta\varepsilon_i}\right)\right]$$
$$= \alpha N + \beta E - \sum_i g_i\, ln\left(1-e^{-\alpha-\beta\varepsilon_i}\right). \qquad (51)$$

From Eq. (51) we write

$$\frac{S}{k} = \beta\left[\frac{\alpha}{\beta}N + E - \frac{1}{\beta}\sum_i g_i\, ln\left(1-e^{-\alpha-\beta\varepsilon_i}\right)\right]. \qquad (52)$$

Equation (52) in conjunction with the second law of thermodynamics gives $\beta = 1/kT$ and $\alpha = -\mu/kT$, where $\mu$ stands for the chemical potential defined as $\mu = \left(\frac{\partial E}{\partial N}\right)_{V,T}$.

As we noted the result in Eq. (50) is independent of the manner in which the energy levels of the particles are grouped into cells so long as the number of levels in each cell is sufficiently large. In view of this, we use the value of $\alpha$ and $\beta$ in Eq. (50) and get the mean occupation number of the level $\varepsilon$ in the form

$$\langle n_\varepsilon \rangle = \frac{1}{e^{(\varepsilon-\mu)/kT}-1} \quad . \qquad (53)$$

From Eq. (53) it is clear that for the mean occupation number to be positive, $\mu < \varepsilon$ for all values of $\varepsilon$. When $\mu$ becomes equal to the lowest value of $\varepsilon$, say $\varepsilon_0$, the occupancy of that particular level becomes infinitely high. This implies that all particles in the gaseous states can go to the lowest energy state. For temperatures at which

$$e^{(\varepsilon-\mu)/kT} \gg 1 \quad . \qquad (54)$$

The gaseous particles obey the Maxwell-Boltzmann statistics given by

$$\langle n_\varepsilon \rangle_{M,B} = e^{(\mu-\varepsilon)/kT}. \qquad (55)$$

Equation (54) tells us that the chemical potential of the system must be negative. Thus, the fugacity $z = e^{\mu/kT}$ of the system must be smaller than unity. The quantity $z$ reflects the tendency of a substance to prefer one phase to another and can literally be defined as the tendency to escape. Moreover, in a quantum mechanical theory $z$ is related to $N$ and $V$ by

$$\frac{N}{V} = z\frac{(2\pi mkT)^{3/2}}{h^3}. \qquad (56)$$

In terms of the de Broglie wavelength

$$\lambda = \frac{h}{\sqrt{2\pi mkT}} \qquad (57)$$

Eq. (56) can be written as

$$\frac{N}{V} = \frac{z}{\lambda^3}. \qquad (58)$$

From Eq. (58), $z$ to be less than unity we must have

$$\frac{\lambda^3 n}{V} \ll 1. \qquad (59)$$

The quantity $n\lambda^3$ ($n = N/V$) is an appropriate parameter in terms of which the various physical properties of the system can be addressed. For example, one can consider three cases ($i$) $n\lambda^3 \longrightarrow 0$: In this case $\lambda \longrightarrow 0$ such that the particle aspect of the gas molecules or atomates over the wave aspect. Obviously, the system is classical. ($ii$) $1 > n\lambda^3 >$: We can expand all physical quantities as a power series in that parameter and investigate how the system tends to exhibit non-classical or quantum behavior. ($iii$) $n\lambda^3 \approx 1$: The system becomes significantly different from the classical one and the typical quantum effects dominate.

From Eq. (57) we write

$$n\lambda^3 = \frac{nh^3}{(2\pi mkT)^{3/2}}. \qquad (60)$$

This expression clearly shows that the system is more likely to display quantum behavior when it is at a relatively low temperature or has a relatively high density of particles. Moreover, for smaller particle mass, the quantum behavior will be more prominent. From Eq. (53) the total number of particles $N$ in the system is obtained as

$$N = \sum_\varepsilon \langle n_\varepsilon \rangle = \sum_\varepsilon \frac{1}{z^{-1}e^{\beta\varepsilon}-1}. \qquad (61)$$

For a large volume $V$, the spectrum of the single-particle state is almost continuous such that the summation on the right side of Eq.(61) can be replaced by integration. The density of states in the neighborhood of $\varepsilon$ is given by

$$\rho(\varepsilon)d\varepsilon = \frac{2\pi V}{h^3}(2m)^{3/2}\varepsilon^{1/2}d\varepsilon \qquad (62)$$

so that

$$\frac{N}{V} = \frac{2\pi}{h^3}(2m)^{3/2}\int_0^\infty \frac{\varepsilon^{1/2}d\varepsilon}{z^{-1}e^{\beta\varepsilon}-1} + \frac{1}{V}\frac{z}{1-z}. \qquad (63)$$

In writing Eq. (63) we have separated out the $\varepsilon = 0$ term in Eq.(61) which has a statistical weight equal to one. Denoting $\frac{z}{1-z}$ by $N_0$ we write Eq. (63) in the form

$$\frac{N-N_0}{V} = \frac{2\pi}{h^3}(2\pi mkT)^{3/2}\int_0^\infty \frac{x^{1/2}dx}{z^{-1}e^x-1} \qquad (64)$$

with $x = \beta\varepsilon$. In terms of Bose-Einstein functions

$$b_\nu(z) = \frac{1}{\Gamma(\nu)}\int_0^\infty \frac{x^{\nu-1}dx}{z^{-1}e^x-1} = z + \frac{z^2}{2^\nu} + \frac{z^3}{3^\nu} + \ldots\ldots \qquad (65)$$

The result in Eq. (64) can be written as

$$\frac{N-N_0}{V} = \frac{1}{\lambda^3}b_{3/2}(z). \qquad (66)$$

The quantity $(N-N_0)$ denotes the number of particles $(N_e)$ in the excited states. Therefore,

$$N_e = V\left(\frac{2\pi mkT}{h^2}\right)^{3/2}b_{3/2}(z). \qquad (67)$$

The function $b_{3/2}(z)$ increases monotonically and is bounded with the largest value

$$b_{\frac{3}{2}} = 1 + \frac{1}{2^{\frac{3}{2}}} + \frac{1}{3^{\frac{3}{2}}} + \cdots\ldots\ldots\ldots$$
$$= \zeta(3/2) = 2.812. \qquad (68)$$

Hence, for all $z$ of interest

$$b_{3/2}(z) \le \zeta(3/2). \qquad (69)$$

In view of Eq. (69) $N_e$ in Eq. (67) will satisfy the condition

$$N_e \le V\left(\frac{2\pi mkT}{h^2}\right)\xi(3/2). \qquad (70)$$

The quantity in Eq. (70) gives the maximum number of particles in the excited states. If the actual number of particles $N$ of the system exceeds this limiting value, then $N_0$ number of particles given by

$$N_0 = N - V \left(\frac{2\pi m k T}{h^2}\right)^{3/2} \zeta(3/2) \qquad (71)$$

will be pushed into the ground state. Since $N_0 = z/(z-1)$, the precise value of $z$ can be determined using

$$z = \frac{N_0}{N_0 - 1} \approx 1. \qquad (72)$$

For $z$ to be one, the chemical potential $\mu$ must be zero. Thus from Eq. (53),$\langle n_e \rangle = \frac{1}{e^{\epsilon/KT} - 1}$. This result shows that for large $N$, there is no limit to the number of particles that can go onto the ground state $\varepsilon = 0$. This curious phenomenon of a macroscopically large number of particles accumulating in a single particle state $\varepsilon = 0$ is referred to as Bose-Einstein condensation. It is purely of quantum mechanical origin and takes place in the momentum space.

The condition for the onset of Bose-Einstein condensation is

$$N > N_e \qquad (73)$$

which gives a critical value of temperature

$$T_c = \frac{h^2}{2\pi m k} \left(\frac{N}{V\zeta(3/2)}\right)^{2/3}. \qquad (74)$$

For given values of $N$ and $V$ Bose-Einstein condensation takes place when temperature $T$ of the gas is less than $T_c$.

**(b) Physical picture of condensate formation**

An atom of mass $m$ at temperature $T$ can be regarded as a quantum mechanical wave packet that has spatial extension of the thermal de Broglie wave length $\lambda = /\sqrt{2\pi m k T}$ given in Eq. (57). From this expression for $\lambda$ one can study the physical changes that occur in the ideal gas as one gradually lowers the temperature. (*i*) As long as the temperature is high, the wave packet is very small such that we can use the classical concept for the trajectory of the wave packet. At such temperatures we imagine atoms as billiard balls that move in the container and occasionally collide. Atoms are distinguishable. This is shown in figure 4(*a*). (*ii*) As the temperature is lowered the wave length increases and the wave aspect of atoms tends to compete with the particle aspect. This is shown in figure 4(*b*). (*iii*) At $T = T_c$ given in Eq. (74) the individual wave packets overlap and we have identity crisis.

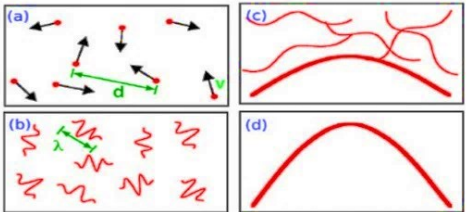

**Figure 4:** Physical changes of atoms during the formation of Bose-Einstein condensation.

The wave packets no longer follow the classical trajectories. At that point indistinguishability becomes important and we need quantum statistics. When quantum indistinguishability dominates, there is a transition to a new phase of matter. The particles come together in a single state and they behave as one big matter wave as shown in figure 4 (c). This is the onset of Bose-Einstein condensation. (*iv*) At T = 0, we get a pure

Bose condensate or a giant matter wave. This is shown in figure 4(d).

It is now clear that the phenomenon of Bose-Einstein condensation, as demonstrated by Einstein, is a consequence of quantum statistics associated with indistinguishability of particles. Naturally, in this context, a very important question arose: What kind of particles obeys Bose statistics and is likely to undergo a phase transition leading to Bose-Einstein condensation? Immediately, after the demonstration of BEC by Einstein, Pauli exclusion principle was formulated. Only after one year the Fermi-Dirac statistics was proposed. Pauli and Dirac thought that all massive particles in the world obey Fermi-Dirac statistics and are fermions. If this was true, Bose-Einstein condensation would never be observed. This was a remark made by Pauli. Meanwhile, Dirac remarked that photons which obey Bose statistics have symmetric wave functions. Thus, particles which are likely to undergo Bose-Einstein condensation at $T < T_c$ must have symmetric wave functions. Understandably, elementary particles (not the carrier of energy) cannot have symmetric wave functions since they are fermions. But there is no bar for composite particles like atoms to undergo Bose-Einstein condensation provided these atoms have integral spins. The total spin of a Bose particle must be an integer, and therefore a boson made up of fermions must contain an even number of them. Neutral atoms contain equal numbers of electrons and protons, and therefore the statistics that an atom obeys is determined solely by the number of neutrons: if N is even, the atom is a boson, and if it is odd, a fermion. Since the alkalis have odd atomic number Z, boson alkali atoms have odd mass numbers A. Likewise for atoms with even Z, bosonic isotopes have even A. In **Table I** we list N, Z and the nuclear spin quantum number I for some alkali atoms and hydrogen.

**Table I.** The proton number Z, the neutron number N, the nuclear spin I.

| Isotope | Z | N | I |
|---------|-----|-----|-----|
| $^1$H | 1 | 0 | ½ |
| $^6$Li | 3 | 3 | 1 |
| $^7$Li | 3 | 4 | 3/2 |
| $^{23}$Na | 11 | 12 | 3/2 |
| $^{39}$K | 19 | 20 | 3/2 |
| $^{40}$K | 19 | 21 | 4 |
| $^{41}$K | 19 | 22 | 3/2 |
| $^{85}$Rb | 37 | 48 | 5/2 |
| $^{87}$Rb | 37 | 50 | 3/2 |
| $^{133}$Cs | 55 | 78 | 7/2 |

To date, most experiments on Bose-Einstein condensation have been made with states having total electronic spin 1/2. The majority of these have been made with atoms having nuclear spin I = 3/2 ($^{87}Rb$, $^{23}Na$, and $^7Li$), while others have involved I = 1/2 (H) and I = 5/2 ($^{85}Rb$). In addition, Bose-Einstein condensation has been achieved for four species with other values of the electronic spin, and nuclear spin I = 0: $^4He^*$ ($^4He$ atoms in the lowest electronic triplet state, which is metastable) which has S = 1, $^{170}Yb$ and $^{174}Yb$ (S = 0), and $^{52}Cr$ (S = 3).

The ground-state electronic structure of alkali atoms is simple: all electrons but one occupies closed shells, and the remaining one is in s-orbital in a higher shell. In **Table II** we list the ground-state electronic configurations for alkali atoms. The nuclear spin is coupled to the electronic spin by the hyperfine interaction. Since the electrons have no orbital angular momentum ($L = 0$), there is no magnetic field at the nucleus due to the orbital motion, and the coupling arises solely due to the magnetic field produced by the electronic spin.

**Table II.** The electron configuration and electronic spin for selected isotopes of alkali atoms and hydrogen.

| Element | Z | Electronic Spin | Electronic Configuration |
|---------|-----|-----|--------------------------|
| H | 1 | 1/2 | $1s^1$ |
| Li | 3 | 1/2 | $1s^2 2s^1$ |
| N$a$ | 11 | 1/2 | $1s^2 2s^2 p^6 3s^1$ |
| K | 19 | 1/2 | $1s^2 2s^2 p^6 3s^2 3p^6 4s^1$ |
| R$b$ | 37 | 1/2 | (Ar) $3d^{10} 4s^2 4p^6 5s^1$ |
| Cs | 55 | 1/2 | (Kr) $4d^{10} 5s^2 5p^6 6s^1$ |

### (c) Steps towards experimental realization

It is now appropriate to ask the question: What are the requirements for observing BEC in the laboratory? We have pointed out that alkali metal atoms are bosons. Thus, any experiment for observation of BEC should start with a gas of alkali metal atoms at the room temperature. The gaseous system should be precooled, trapped and cooled to a temperature preferably below the critical temperature and then imaged to get the signature of BEC. In fact, the first three experiments on BEC used dilute atomic gases of rubidium [8], lithium [9] and sodium [10]. It is true that BEC was first observed in these three experiments. However, it appears that superfluidity in helium was considered by London as early as 1938 as a possible manifestation of BEC. However, evidence for BEC in helium was found much later from the analysis of momentum distribution of the atoms measured in neutron-scattering experiment [25]. On the other hand, in a series of experiments hydrogen atoms were first cooled in a dilute refrigerator, then trapped by magnetic field and further cooled by evaporation. This approach has come very close to observing BEC. The main problem in observing BEC in this system comes from the fact that the hydrogen atoms, rather than being in atomic state, form molecules [26].

In the 1980's laser-based techniques were developed to trap and cool neutral atoms [27]. Technically, trapping and cooling in this approach go by the names magnetooptical trapping and laser cooling. Alkali metal atoms are well suited to laser-based methods because their optical transitions can be excited by available lasers and because they have a favorable internal energy-level structure for cooling to very low temperatures. Once they are trapped, their temperature can be lowered further by evaporative cooling. Let us first briefly outline what are the effects of trapping on the atomic system and how atoms are trapped.

### (i) The effects of trapping

The number of atoms that can be put into the trap is not truly macroscopic such that the thermodynamic limit is never achieved. We, therefore, begin by considering the effect of finite particle numbers on $T_c$, the critical temperature for the onset of Bose-Einstein condensation. The expression for $T_c$ in Eq. (74) refers to $N$ (very large) number of particles confined in a three-dimensional box. If instead we consider the atoms to be confined in a three-dimensional harmonic well, the expression for $T_c$ modifies to

$$T_c = \frac{\hbar \bar{\omega} N^{1/3}}{[\zeta(3)]^{1/3}}, \qquad (75)$$

where $\bar{\omega} = (\omega_1 \omega_2 \omega_3)^{1/3}$, $\omega$ be the classical oscillator frequency. In this context we note that in most cases the confining traps are well approximated by harmonic potentials. When the number of particles is extremely high, we can neglect the zero-point energy in the harmonic trap. This is, however, not true when the system consists of finite number of atoms. The finiteness of the number of particles calls for zero-point energy to be taken into account. This reduces the critical temperature by an amount $\Delta T_c$ such that

$$\frac{\Delta T_c}{T_c} = -\frac{\zeta(2)}{2[\zeta(3)]^{2/3}} \frac{\omega_m}{\bar{\omega}} N^{-1/3}. \qquad (76)$$

where $\omega_m = (\omega_1 + \omega_2 + \omega_3)/3$. Clearly, from Eq. (76) $\frac{\Delta T_c}{T_c} -> 0$, as $N -> \infty$. Thus, we see that one of the effects of trapping is to lower the critical temperature by confining a finite number of atoms. Besides finiteness of the system trapping makes the Bose gas inhomogeneous such that density variation occurs on a characteristic length scale, $a_{ho} = \sqrt{\hbar/(m\omega_m)}$, provided by the frequency of the trapping oscillator. This is a major difference with respect to other systems like the super fluid helium where the effects of inhomogeneity take place on a microscopic scale in the coordinate space. Inhomogeneity of super-fluid helium, in fact cannot be detected in the coordinate space such that all observations are made in the momentum space. As opposed to this, the inhomogeneity of the Bose gas is such that both coordinate and momentum spaces are equally suitable for observations.

In the above we talked about harmonic confinement. Physically such confinements are achieved by applying appropriately chosen inhomogeneous magnetic fields, often called magnetic trap. Magnetic traps are used to confine precooled gaseous system. We shall first discuss the method of precooling and then talk about the principle of magnetic trapping.

### (ii) Method of precooling

Laser beams are often used to precool the atomic vapour and the method used goes by the name laser cooling. The physical mechanism by which the collision between photons and atoms reduces the temperature of the atomic vapour can be visualized as follows.

If an atom travels toward the laser beam and absorbs a photon from the laser it will be slowed down by the photon impact. Understandably, totality of such events will lower the temperature. On the other hand, if the atom moves away from the photon, the latter will speed up resulting in the increase of temperature. Thus, it is necessary to have more absorptions from head on photons if our goal is to slow down the atoms with a view to lower the temperature. One simple way to accomplish this in practice is to tune the laser slightly below the resonance absorption of the atom.

Suppose that the laser beam is propagating in a definite direction. An atom in the gaseous system can move towards the beam or it may move away from the beam. In both cases the frequency of the photon will be Doppler shifted. In the first case the frequency of the laser beam will increase while in the other case the frequency will be decreased. In the case of head on collision the photon will be absorbed by the atom via resonance only when the original laser beam is kept below the frequency of atomic resonance absorption. When the atom and photon travel in the opposite direction there cannot be momentum transfer from the photon to the atom because Doppler shift in this case produces further detuning of the already detuned laser beam.

The explanation presented above provides only a simple-minded realization of laser cooling. The physics of any typical experiment is much more complicated than that because the absorption of photon by atom is also accompanied by an emission process. The emission and absorption produce a velocity dependent force that is responsible for cooling. A technique of laser cooling based on velocity-dependent absorption process goes by the name Doppler cooling. Doppler cooling can also be used in an arrangement called optical molasses where cooling is done in all three-dimensions. There is still another variant of laser cooling that goes by the name Sisyphus cooling. The mechanism of Sisyphus cooling is somewhat sophisticated. It involves a polarization gradient generated by two counter propagating linearly polarized laser beams with perpendicular polarization directions.

**(iii) Basic principles of magnetic trapping**

Magnetic traps are used to confine low temperature atoms produced by laser cooling. These traps use the same principle as that in the Stern-Gerlach experiment. Otto Stern and Walter Gerlach used the force produced by a strong inhomogeneous magnetic field to separate the spin states in a thermal atomic beam as it passes through the magnetic field. But for cold atoms the force produced by a system of magnetic coils bends the trajectories right around so that low energy atoms remain within a small region close to centre of the trap. This can be realized as follows.

A magnetic dipole moment $\vec{\mu}$ in a magnetic field $\vec{B}$ has energy

$$V = -\vec{\mu}.\vec{B}. \tag{77}$$

For an atom in a hyperfine state $|IJFM_F\rangle$, $V$ corresponds to a Zeeman energy

$$V = g_F \mu_B M_F B, \tag{78}$$

where $\mu_B =$ Bohr magneton and

$$g_F \simeq g_J \frac{F(F+1)+J(J+1)-S(S+1)}{2F(F+1)}. \tag{79}$$

The magnetic force acting along $z$ - direction

$$P = -g_F \mu_B M_F \frac{dB}{dz}. \tag{80}$$

We shall now make use of Eqs. (77) and (78) to indicate (*i*) why precooling is necessary for the use of magnetic trap? and (*ii*) what should be the nature of the magnetic field that produces a trap useful for confining BEC? From Eq. (77) the energy depth of the magnetic trap is determined by $\mu_i B$. The atomic magnetic moment $\mu_i$ is of the order of Bohr magneton $\mu_B$ which in temperature units $\approx 0.67$ Kelvin/Tesla. Since laboratory magnetic fields are generally considerably less than 1 Tesla, the depth of magnetic traps is much less than a Kelvin, and therefore atoms must be cooled in order to be trapped magnetically.

For confinement, Zeeman energy must have a minimum. We can consider two different cases for Eq. (78). Case 1: $M_F g_F > 0$. Here the Zeeman energy can be minimum if $B$ has a local minimum. Case 2: $M_F g_F < 0$. In this case $V$ can have a local minimum if $B$ has a local maximum. Maxwell's equations do not allow a maximum of a static field. As a result, the trapping of atoms for $M_F g_F < 0$ is not allowed. In view of the above one can trap atoms only in a minimum of a static magnetic field.

**(iv) More details for magnetic trapping of neutral atoms**

We have noted the following.

Confinement of neutral atoms depends on the interaction between an inhomogeneous magnetic field and atomic multipole moment. (Dipoles may be trapped by the local field minimum. Field configurations with a minimum in $|\vec{B}|$ may be divided into two classes: (a) where the minimum of the field is zero, and (b) where it is non-zero. The original quadrupole trap as devised by Paul in NIST or the so-called Paul trap is shown in figure 5. It belongs to class (a). This trap consists of two identical coils carrying opposite currents and has a single center where the field is zero. It is the simplest of all possible magnetic traps. When the coils are separated by 1.25 times their radius, such a trap has equal depth in the radial (x-y) plane and longitudinal (z - axis) directions. Its experimental simplicity makes it most attractive, both because of construction and of optical access to the interior.

The quadrupole trap suffers from an important disadvantage. The atoms assemble near the center where $B \approx 0$. As a result, Zeeman sublevels ($|IJFM_F\rangle$) have very small energy separation. The states with different magnetic quantum numbers mix together and atoms can make transition from one value of $M_F$ to another due to fluctuation in the field. These nonadiabatic transitions allow the atoms to escape and reduce the lifetime of atoms in the trap. There have been two major efforts to circumvent the disadvantage of using the simple quadrupole trap. In the first case one superimposes an oscillating biased magnetic field on the quadrupole trap. Admittedly, this makes the magnetic field $\vec{B}$ time dependent. The time average of the resulting field remains nonvanishing at the center. An alternative approach, adapted by the MIT group of Ketterle, is to apply a laser field in the region of the node in the magnetic field.

Instead of using traps having a node in the magnetic field, one can remove the hole by working with magnetic field configurations that have a nonzero field at the minimum. The schematic diagram of such a magnetic field configuration is given in figure 6. Here four parallel wires arranged at the corner of a square produces a quadrupole magnetic field when currents in adjacent wires flow in the opposite directions. The resulting radiation forces repel atoms from the vicinity of the node, thereby reducing losses. This field has a linear dependence on the radial coordinate $r$ and is given by

$$|\vec{B}| = b'r, \qquad (81)$$

where $b' = \frac{\partial B_x}{\partial x} = -\frac{\partial B_y}{\partial y}$ obtained from $\vec{\nabla}.\vec{B} = 0$. Using Eqs. (78) and (79)

$$V(r) = g_F \mu_B M_F b'r. \qquad (82)$$

Variation of the quantity $|\vec{B}|$ in Eq. (77) for the quadrupole trap is shown in figure 6(b). Clearly, $|\vec{B}| = 0$ at $r = 0$. In the so-called Ioffe trap this problem is circumvented by using two circular coils which enclose the parallel wires as shown in figure 7(a). In both coils current flow in the same direction. The magnetic field for this configuration if given by

$$|\vec{B}| \approx B_0 + \frac{b'^2 r^2}{2B_0} \qquad (83)$$

where $B_0$ is a magnetic field in the $z$ direction produced by currents in the circular coils. For $|\vec{B}|$ in (83) a plot similar to that in figure 6(b) looks like the plot in figure 7(b). Clearly, this field has a nonzero value at $r = 0$.

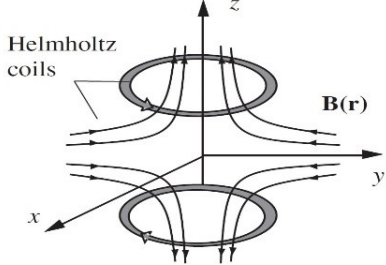

**Figure 5:** Diagram of Paul trap.

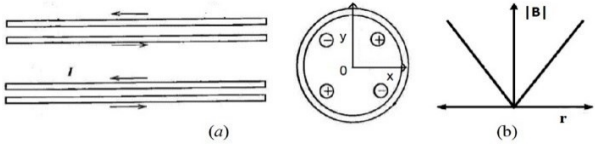

**Figure 6 :** (a) Linear quadrupole trap and (b) Magnetic field with radial coordinate $r$.

### (v) Optical trapping

Magnetic traps provide an efficient method to confine cold atoms. The basic principle of optical trapping is as follows. The interaction between an atom and the electric field is given by

$$H' = -\vec{d}.\vec{\varepsilon}, \qquad (84)$$

Where $\vec{d}$, electric dipole moment and $\vec{\varepsilon}$, the electric field vector. Perturbation $H'$ changes the ground-state energy by

$$\Delta E_g = -\frac{1}{2}\alpha\varepsilon^2, \qquad (85)$$

where $\alpha =$ static atomic polarizability. Expression in Eq. (85) refers to an energy change produced by a static electric field. The electric field in laser light is time-dependent. For a time-dependent electric field the expression in Eq. (85) modifies to

$$\Delta E_g = -\frac{1}{2}\alpha(\omega)\langle\varepsilon(\vec{r}, t)^2\rangle, \qquad (86)$$

where $\alpha(\omega)$ is the frequency-dependent polarizability.

An atom excited by the electric field is likely to decay by spontaneous emission. If this fact is taken into account, the frequency dependent polarizability becomes a complex quantity such that $\Delta E_g$ (86) could be written as

$$\Delta E_g = V_g - i\hbar\Gamma_g/2, \qquad (87)$$

where the real part $V_g$ corresponds to a shift in energy of the ground state while the imaginary part represents the finite lifetime $1/\Gamma_g$ of the ground state due to the transition to the excited state induced by the radiation. In more detail,

$$\alpha(\omega) \approx \frac{|\langle e|\vec{d}.\hat{\varepsilon}|g\rangle|}{E_e - i\hbar\Gamma_e/2 - E_g - \hbar\omega}. \qquad (88)$$

Here $1/\Gamma_e =$ lifetime of the excited state. Using Eq.(88) in Eq.(86) and comparing the result with (50) we get

$$V_g = -\frac{1}{2}\alpha_R(\omega)\langle\varepsilon(\vec{r}, t)^2\rangle_t \qquad (89)$$

with the real part of $\alpha(\omega)$

$$\alpha_R(\omega) = \frac{(\omega_{eg} - \omega)|\langle e|\vec{d}.\hat{\varepsilon}|g\rangle|^2}{\hbar[(\omega_{eg} - \omega)^2 + (\Gamma_e/2)^2]}, \qquad (90)$$

where $\omega_{eg} = (E_e - E_g)/\hbar$. The force corresponding to the potential in Eq. (89) is given by

$$F_{dipole} = \frac{1}{2}\alpha_R(\omega)\vec{\nabla}\langle\varepsilon(\vec{r}, t)^2\rangle_t. \qquad (91)$$

From (90) we see that if $\omega > \omega_{eg}$, $\alpha_R(\omega)$ is negative and if $\omega < \omega_{eg}$, $\alpha_R(\omega)$ is positive. In the first case the laser beam is called blue detuned while in the second case we have red detuning. For red detuning the force acts along the higher field. On the other hand, for blue detuning the force acts along lower field. By focusing a laser beam it is possible to create a radiation field whose intensity has a maximum in space. If the frequency of the light is detuned to the red, the energy of the ground-state atom has a spatial minimum, and therefore it is possible to trap atoms.

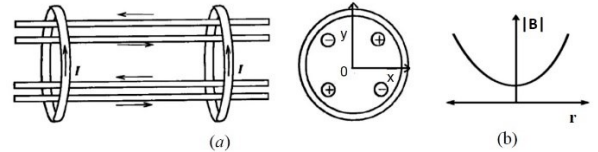

**Figure 7:** (a) Ioffe trap: combination of a linear magnetic quadrupole and an axial biased field. (b) Magnetic field that provides radial confinement of atoms.

### (vi) Evaporative cooling

The temperature reached by laser cooling is quite low, but not low enough to produce Bose-Einstein condensation in gases at densities that are realizable experimentally. A very effective technique of reducing the temperature of the magnetically trapped laser cooled atoms goes by the name evaporative cooling. In the experiment performed to date, Bose-Einstein condensation of alkali gases is achieved by using evaporative cooling. The basic physical effect in evaporative cooling is that, if particles escaping from a system have an energy higher than the average energy of particles in the system, the remaining particles are cooled. Evaporative cooling could be carried out by lowering the strength of the trap. But this reduces the density and eventually makes the trap too weak to support atoms against gravity. However, this method has been successfully used for $Rb$ and $Cs$ atoms in dipole-force traps. There is another important method for evaporative cooling. Here precisely controlled evaporation is carried out by using radio frequency radiation that changes the spin state of an atom from a low field seeking one to a high field seeking one, hereby expelling atoms from the trap.

### (d) Observing the BEC in the laboratory
In a BEC the observable quantity is the density profile. There are two important methods to observe the density profile. The first one is called absorptive imaging while the second one goes by the name phase-contrast imaging.

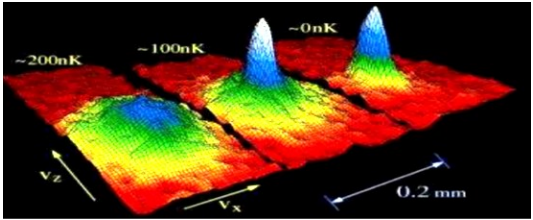

**Figure 8:** Velocity distribution of $^{87}$Rb condensates where red colour represents the region having lower number of atoms while white colour represents region of larger number of atoms [10].

### (i) Absorptive imaging:
Light at a resonant frequency for the atom will be absorbed on passing through an atomic cloud. Thus, measuring the absorption profile one can obtain information about the density distribution. The spatial resolution can be improved by allowing the cloud to expand before measuring the absorptive image. A drawback of this method is that it is destructive, since absorption of light changes the internal states of atoms and heats the cloud. An observation of Bose-Einstein condensation by absorption imaging is displayed in figure 8. It shows absorption vs. two spatial dimensions. The Bose-Einstein condensate is characterized by its slow expansion observed 6 ms after the atom trap was turned off. The left picture shows an expanding cloud cooled just above the critical temperature; middle: just after the condensate appeared; right: after further evaporative cooling has left an almost pure condensate.

### (ii) Phase-contrast imaging:
This method exploits the fact that the refractive index of a gas depends on its density. Therefore, the optical path length is changed by the medium. Here a light beam is passed through the cloud. This is allowed to interfere with a reference beam that has been phase shifted. The change in optical path length as evident from the interference pattern is then converted into intensity variation for observation.

## VI. CONDENSED ATOMIC GASES
### (a) A new platform for studying condensed-matter physics
Dilute atomic gases in BECs are distinguished from the condensed-matter systems by the absence of strong and complex interaction. Despite that, studies in these quantum gases have become an interdisciplinary field of atomic and condensed matter physics. Topics of many-body physics can now be studied with the methods of atomic physics. In this way, BEC provides a new subfield that can confidently be used to simulate properties condense-matter systems [28].

The observation of the condensate's density distribution can be regarded as a direct visualization of the microscopic wave function. The time evolution of the squared wave function of a single condensate has been recorded non-destructively in real time [29, 30]. In a recent article Hannaford and Sacha [31] reviewed the case of a BEC of ultracold atoms bouncing resonantly on an oscillating atom mirror such that the period of the bouncing atoms is equal to an integer multiple of the period of the driving mirror. The bouncing BEC can exhibit dramatic breaking of time-translation symmetry, allowing the creation of discrete time crystals.

### (b) Problem to inject spin-orbit coupling in BECs
The ultracold atoms which form the condensates are electrically neutral. Consequently, the intrinsic physics relevant to the charge degrees of freedom is absent in these systems. Ever since Bose-Einstein condensation (BEC) was achieved in atomic gases, one of the main tasks for both theory and experiment has been to introduce charge physics into neutral atoms creating gauge fields by artificial means. Experimentally, the first artificial magnetic field was synthesized in a harmonically trapped BEC through the rotation of the external trapping potential [32]. In the rotating frame, this leads to a Lorentz force and an antitrapping potential for atoms, where the amplitude and frequency of the anti-trap are proportional to the rotational frequency. However, this technique is limited because the strength of the anti-trap cannot exceed that of the trapping potential, which implies that it cannot be used for spatially homogeneous BECs. In this case, the magnetic field has to be generated artificially.

### (c) Way to artificial gauge field
In the last few years there have been several attempts [33] to realize artificial gauge fields for quantum gases and thus overcome the problem arising due to neutrality of atoms to introduce spin-orbit interaction in an atomic cloud. One of these schemes relies on the notion of geometric phase [34] acquired by a quantum mechanical wave function while evolving in a parameter space of the Hamiltonian. The phase angle is defined in terms of an integral over a vector valued function often called the Berry connection. The Berry connection corresponds to an artificial vector potential for neutral atoms. To implement this idea the researchers from NIST and from University of Maryland exploited the space-dependent coupling of the atoms with a properly designed configuration of laser beams. The synthetic gauge field arises when the system adiabatically follows one of the local eigenstates of the light-atom interaction Hamiltonian [35]. Since 2009 several experiments have been successful in realizing ultra-cold atomic gases coupled to artificial gauge fields. For instance, a space dependent atom-light coupling was employed to generate an effective magnetic field to exert a Lorentz-like force on neutral bosons [36]. This procedure has also been used to generate quantized vortices in BECs.

### (d) Spin-orbit coupled BECs
After engineering gauge fields in BEC the same group of workers attempted to simulate the coupling between an atom's spin and its motion with a view to realize the effect of spin-orbit coupling (SOC) in ultra-cold atomic systems. In atomic physics, SOC is an interaction between the electron's spin and its motion about the nucleus. For solids, SOC provides a link between the electron's spin and its motion in the crystal lattice. In both cases the SOC arises due to interaction of electrons with the electric fields that exist inside atoms and solids. But there is no charge field for atoms moving in a BEC. Consequently, Spielman and his group [37] sought a way to link the internal spin of an atom to its momentum. They had chosen to work with a BEC of $^{87}Rb$ atoms and focused their attention on the atom's electronic ground state, $5S_{1/2} F=1$. In a typical experiment, a degenerate cloud of $^{87}Rb$ is prepared in a crossed optical dipole trap.

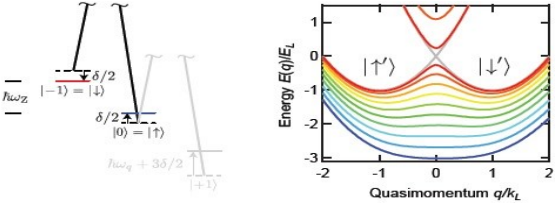

**Figure 9:** Level diagram. Two Raman lasers couple the two states $| F = 1, m_f = -1 >$ and $|F = 1, m_f = 0 >$ of the F = 1 hyperfine manifold of $^{87}$Rb, which differ in energy by a Zeeman splitting. The lasers have frequency difference $\Delta \omega_L = \omega_Z + \delta$, where $\delta$ is a small detuning from the Raman resonance. The state $| F = 1, m_f = +1 >$ can be neglected since it has a much larger detuning, due to the quadratic Zeeman effect.

By applying a homogeneous magnetic bias field along the $x$-direction (as shown in figure 9), the $F=1$ ground state is split into three energy levels, $m_f = 0, \pm 1$. Two Raman lasers whose projections on the $x$-direction point oppositely are used to couple the energy levels $|F = 1, m_f = -1>$ and $|F = 1, m_f = 0>$, and $|F = 1, m_f = 0>$ and $|F = 1, m_f = 1>$. These energy levels can be interpreted as pseudo-spins. However, to simulate spin ½ systems, the third energy level have to be moved out of resonance. This can be done by applying a large magnetic field so that quadratic Zeeman effect shifts the energy splitting between $|F=1, m_f = -1>$ and $|F=1, m_f = 0>$ to a larger value than that of the $|F = 1, m_f = 0>$ and $|F = 1, m_f = 1>$. Choosing an appropriate frequency between the Raman Lasers then allows one to address only the $|F = 1, m_f = -1>$ and $|F = 1, m_f = 0>$ transitions. This is how, by dressing two spin states with a pair of lasers, Spielman et al could engineer SOC with equal amount of Rashba [38] and Dresselhaus [39] coupling in a neutral atomic Bose-Einstein condensate. The synthetic spin-orbit coupling for neutral atoms was subsequently realized in other laboratories [40-43]. Thus, it became quite urgent to investigate theoretically how does the SOC affect the properties of usual BECs (without spin-orbit coupling) [44-46].

### (e) Effect of spin-orbit coupling on the condensates

The behaviour of Bose-Einstein condensates is studied in the first-order of approximation by using Gross-Pitaevskii equation (GPE) [47,48] which is, in fact, the well-known Schrödinger equation plus a nonlinear term that takes into account of the interaction between atoms forming the condensates. The self-trapped localized states can be considered both in one-dimensional or higher-dimensional geometries. In multidimensional settings, stability of these states becomes a major issue due to critical and supercritical collapse [49, 50]. As opposed to this, in the one-dimensional case we have stable soliton states in diverse systems [51]. For a one-dimensional model of the BEC we assume that we have a dilute Bose gas and the radial trapping frequency of the condensate is much larger than the axial frequency, i. e. the BEC is cigar shaped. In this case the GPE is given by

$$i\hbar \frac{\partial \psi}{\partial t} = \left(-\frac{\hbar^2}{2m}\partial_x^2 + V_{tr}(x) - g|\psi|^2\right)\psi, \quad (92)$$

where $m$ stands for the mass of atoms in the condensate and $\psi$ for the complex order parameter or wave function of the condensate. The quantities $V_{tr}$ and $g$ represent the trapping potential and coupling constant of the atom-atom interaction. It is well known that for attractive inter-atomic interaction ($g > 0$) Eq. (92) supports soliton solution such that we have a highly stable condensate.

Equation (92) governs the dynamics of the BEC in the absence of spin-orbit coupling. A BEC with experimentally realized SOC is characterized by a spinor order parameter $\psi = (\psi_\uparrow, \psi_\downarrow)$ where $\psi_\uparrow$ and $\psi_\downarrow$ are related to the two pseudospin components of the BEC. The dimensionless equations of motion for $\psi_{\uparrow\downarrow}$ can be written as [52]

$$i\partial_t \psi_\uparrow = \left(-\frac{1}{2}\partial_x^2 - ik_L\partial_x + V_{tr}(x) - |\psi_\uparrow|^2 - \beta|\psi_\downarrow|^2\right)\psi_\uparrow + \Omega\psi_\downarrow, \quad (93)$$

$$i\partial_t \psi_\downarrow = \left(-\frac{1}{2}\partial_x^2 + ik_L\partial_x + V_{tr}(x) - \beta|\psi_\uparrow|^2 - \gamma|\psi_\downarrow|^2\right)\psi_\downarrow + \Omega\psi_\uparrow. \quad (94)$$

Here $k_L$ is the wave number of the Raman laser which couples two atomic hyperfine states and $\Omega$, the Raman or Rabi frequency. The quantities $\beta$ and $\gamma$ related to the $s$-wave scattering $\alpha_{ij}(i,j = 1, 2)$. In an interesting paper Xu, Zhang and Wu [53] made use of equations similar to those given above to study the properties of bright solitons ($\alpha_{ij} < 0$) and observed that the stationary bright solitons which are the ground states of the system have nodes in their wave function. For a conventional BEC without SOC its ground state must be nodeless. This is consistent with the so-called 'no node' theorem for the ground state of bosonic system [54]. In fact, the soliton in SOC-BEC is fundamentally different from the conventional one since the spin-orbit coupling breaks the Galilean invariance of the system. This lack of invariance was experimentally demonstrated [55] by studying the dynamics of SOC-BEC loaded in a translating optical lattice.

Optical lattices use standing wave patterns of counter-propagating laser beams to create a periodic potential for ultracold atoms. In most experiments [56-58] optical lattices act as external potentials and thus introduce a state-independent intrinsic periodicity in the system. We shall refer to these type lattices as linear optical lattice (LOL). Besides LOL, it is possible to consider a nonlinear optical lattice (NOL). The latter possess symmetry properties that depend on the wave function representing the state of the system [59]. The BECs loaded in optical lattices (OLs) are known to exhibit many interesting physical phenomena such as Bloch oscillation, Landau-Zener tunnelling, Mott Transition etc. [60]. In view of this, there has been a great deal of activities for studying dynamics of BEC solitons in OLs.

It will be interesting to study the effect of optical lattices on the structure of the bright soliton in the quasi-one-dimensional SOC-BEC. Recently, it has been found that parameters of the lattice potentials can be used to provide useful control over the number of nodes of the bright soliton and thereby make attempts to restore the Galilean invariance [61]. In this context we note that studies on restoration of Galilean invariance of physical systems are of relatively recent origin and have mainly been undertaken for nuclear force problem [62,63]. While looking for control over the number of nodes it has been seen that the soliton with large number of nodes is less stable compared to one having fewer number of nodes. This indicated that the synthetic spin-orbit coupling induces instability in the ordinary matter-wave soliton. In the following we shall study how the synthetic spin-orbit affect the atomic density distribution in the BEC.

As is well known that in atoms the spin-orbit coupling is the interaction between the electron's spin and its orbital motion around the nucleus. When an electron moves in the finite electric field of the nucleus, the spin–orbit coupling causes splitting in the electron's atomic energy levels thus leads to new spectroscopic phenomena. Spin-orbit coupling also lies at the core of condensed matter. In the presence of strong SOC, there can appear a wide variety exotic physical phenomenon in solid-state systems. For example, the coupling between an electron's spin and its momentum is crucial for topological insulators [64, 65] as well as for Majorana fermions [66]. In this context we also note that magnetic fields influencing the motion of electrons in a semiconductor are at the base of quantum Hall effect [67]. The spin electronics also called spintronics is an emerging field of basic and applied research in physics that aims to exploit the role played by electron spin in solid materials with a view to develop semiconductors that can manipulate the magnetic property of an electron [68].

Ultracold atomic gases are good candidates to investigate the above interesting quantum phenomena. In this respect, as already noted, the main difficulty arises from the fact that atoms are neutral particles, and consequently, like electrons, they cannot be coupled to gauge fields so as to exhibit any coupling between their spin and their center of mass motion. But the experimental realization of synthetic spin-orbit coupling [37] has removed this stumbling block and opened the door for simulating many observed phenomena in condensed-matter physics. In this respect a very important problem consists in studying the response of condensates' density profile to changes in the strength of the tuneable synthetic spin-orbit coupling. As regards the density profile of a quasi-one dimensional SOC-BEC we focus our attention on the solution of the coupled Eqs.(93) and (94). From these equations it is clear that the dynamics of SOC-BEC, in addition to the parameters of the trapping potential and strength of the inter-atomic interaction, depends crucially on the spin-orbit coupling parameter $\kappa_L$ and the so-called Rabi frequency $\Omega$. The parameter $\kappa_L$ physically represents the wave number of the Raman laser that inject spin-orbit coupling in the condensate. Depending on the choice for the values of $\kappa_K$ and $\Omega$ one can distinguish two different regions in the linear energy spectrum of the system. In region I, characterized by $\kappa_L^2 < \Omega$ has a single minimum and the associated GPE with attractive atom-atom interaction supports bright soliton solution [69] of the nonlinear Schrodinger equation. On the other hand, in region II with $\kappa_L^2 > \Omega$, the dispersion curve posses two minima say, $\pm k_0$, of the system. Two different solutions corresponding to these minima have also been given in ref. 52. In addition, we can have a linear superposition of these solutions that form a strip phase [40,70]. These wave functions can be used to obtain results for corresponding normalized to unity probability density distributions. Information theoretic measures of such probability densities have been found to provide a useful basis to examine the effect of any perturbation in the system [71,72]. In our case the synthetic SOC perturbs the density distribution of the conventional BEC. We now discuss below how such measures have recently been used [73] to study the perturbative effect of our interest.

Two popular information measures of a normalized to unity probability density $\rho(x)$ are given by the so-called Shannon entropy [74]

$$S_\rho = -\int_{-\infty}^{\infty} \rho(x) \ln \rho(x) dx \qquad (95)$$

and Fisher information [76]

$$I_\rho = \int_{-\infty}^{\infty} \rho(x) \left[\frac{d}{dx} \ln \rho(x)\right]^2 dx. \qquad (96)$$

In the momentum space, results corresponding to the one-dimensional quantities in Eqs. (95) and (96) are given by

$$S_\gamma = -\int_{-\infty}^{\infty} \gamma(p) \ln \gamma(p) dp \qquad (97)$$

and $\quad I_\gamma = \int_{-\infty}^{\infty} \gamma(p) \left[\frac{d}{dp} \ln \gamma(p)\right]^2 dp, \qquad (98)$

The results of the above information theoretic quantities are subject to the constraints

$$S_\rho + S_\gamma \geq 2.14473 \qquad (99)$$

and $\qquad I_\rho I_\gamma \geq 4. \qquad (100)$

The relation in Eq. (99) represents a stronger version of the uncertainty relation introduced by Bialynicki-Birula and Mycielski [76] while that in Eq .(100) is due to Hall [77]. The Fisher based uncertainty relation has been re-derived by Dehesa et. al [78].

The information measures defined in Eqs. (97) and (98) provide complementary descriptions of disorder in the system. From mathematical point of view, one is a convex while the latter is concave [80]. When one grows the other diminishes. The larger values of the position-space Shannon entropy are associated with delocalization of the underlying densities while the smaller values imply localization Understandably, the opposite is true for the Fisher information. In view of this properties of $S$ and $I$ can be gainfully used to investigate how does the density distribution of a quantum many-body system respond to external perturbations.

Thus, numbers for both Shannon entropy and Fisher indicate that by increasing the strength of the spin-orbit coupling constant we go from a delocalized to localized atomic distribution in the condensate. The corresponding momentum -space quantities $S_\gamma(I_\gamma)$ increase(decrease) as $\kappa_L$ become large. The results for position - and momentum - space information measures never violate uncertainty relations in Eqs. (99) and (100). In region II ($\kappa_L^2 > \Omega$), $S_\rho$, $S_\gamma$, $I_\rho$ and $I_\gamma$ as a function of $\kappa_L$ exhibit opposite behavior as observed for these quantities in region I. This establishes that for $\kappa_L^2 > \Omega$ the atomic distribution in the condensate becomes delocalized as we increase the strength of the SOC. This is just opposite to what we observed for the condensate in spectral region I. For the BEC in the stripe phase the results for $S$ and $I$ exhibit similar behavior as that of the condensate in region II i. e. BEC becomes more localized as we increase the strength of the SOC. At the lowest admissible value of $\kappa_L$ under the constraint $\kappa_L^2 > \Omega$ the result for $I_\rho$ has been found to be rather inconsistent. This implies that the strip phase cannot exist in BECs unless the synthetic spin-orbit coupling is strong enough. In general, the numbers for $I_\rho$ are greater than the corresponding values of $I_\rho$ in spectral region II by an order of magnitude. Very high values of $I_\rho$ tend to establish that in the stripe phase the density distribution is highly localized. This is quite expected since this phase characterizes the supersolid properties of the SOC BEC. From the results for $I_\rho$ and $I_\gamma$ it is seen that values of the uncertainty $I_\rho I_\gamma$ are very large for all admissible values of $\kappa_L$. It is, therefore, tempting to infer that supersolidity is a purely quantum mechanical Phenomenon.

## VII. CONCLUDING REMARKS

The present work is a modest attempt to pay homage to the great Indian scientist Satyendra Nath Bose. His pioneering work came from India when the center of intense scientific activity was Europe. We began by noting that Bose's intellectual development was unusual and he was destined to play an inspiring role in the scientific and cultural life of our country. He was well versed in Bengali, English, French, German and Sanskrit. Bose always had wanted science to be taught in mother language and he tried his best to achieve the goal. In recognition of Bose's effort to popularize science through the mother language, poet Rabindra Nath Tagore invited Bose to Santiniketan and dedicated the book 'Visva-Parichaya' to him [80]. This book gives an elementary account of cosmic and microcosmic world in Bengali. We noted in the text that, in honor of S. N. Bose, the British physicist P. A. M. Dirac, the originator of the relativistic electron theory, coined the term 'boson' for particles which obey Bose statistics.

Next, we focused our attention on the statistics introduced by Bose and its application to derive Planck's law of radiation, which opened a new window into the quantum world and made him immortal in the history of science. In fact, Bose read Planck's paper on the distribution of energy in blackbody radiation to teach this material in his class. He was disturbed by the ad hoc assumptions of Planck as were used to derive the law. In 1924, Saha stayed with Bose in Dacca (now called Dhaka) and pointed out the papers of Wolfgang Pauli, Paul Ehrenfest and their relation to Einstein's 1917 paper. This led him to develop the so-called Bose statistics - a new method to count the states of indistinguishable particles - and apply it to his derivation of Planck's law. After the publication of Bose's work by the recommendation of Albert Einstein, Einstein himself extended the work of Bose to material particles. This led to the birth of Bose-Einstein statistics and theoretical prediction of a macroscopic quantum phenomenon - Bose-Einstein condensation (BEC). Since then, there were many attempts to observe BEC in the laboratory. Now we know that BECs are produced by cooling a dilute atomic gas to nano kelvin temperature using laser and evaporative cooling. We have tried to briefly outline the series of events that ultimately led to the experimental realization of BECs.

There has been a growing interest in the physics of cold atoms. For example, the NIST group generated Abelian gauge fields to introduce a synthetic spin - orbit interaction in the electrically neutral cold atoms of the Bose-Einstein condensates. Such interactions have been found to drastically affect properties of the conventional condensates without spin-orbit coupling. We have provided here two examples in respect of this. In the first one we studied the interplay between the spin-degrees of freedom and nonlinear atomic interaction by loading a one-dimensional SOC BEC in an optical lattice. In the second one we made use of two information theoretic measures to critically examine the effect of spin-orbit coupling on the density profiles of a quasi-one-dimensional condensate with attractive inter - atomic interaction. Interestingly enough, it was found that the BEC in the stripe phase exhibits supersolid properties and supersolidity is a purely quantum mechanical phenomenon.

## ACKNOWLEDGEMENT

One of the authors (GAS) would like to acknowledge funding from the 'Science and Engineering Research Board, Govt. of India' through Grant No.CRG/2019/000737.

## REFERENCES

[1] Narlikar, J. V., Lectures on general relativity and cosmology, (Mcmillan Press, London, 1979).

[2] Mehra, J. and Rechenberg, H., The Historical Development of Quantum Theory, (Springer-Verlag New York, Inc. 1982).

[3] A. Pais, Subtle is the Lord: The science and the life of Albert Einstein, (Oxford University Press, USA,1998).

[4] Blanpied, W. A., Satyendranath Bose: Co-Founder of Quantum Statistics, Am. J. Phys., 1972, vol. **40**, p. 1212-1220.

[5] Hau, L. V., Busch, B. D., Liu, C., Burns, M. M. and Golovchenko, J. A., Cold Atoms and Creation of New States of Matter: Bose-Einstein Condensates, Kapitza States, and '2D Magnetic Hydrogen Atoms', arXiv:cond-mat/9804277v1; 24 Apr. (1998).

[6] Chatterjee, S. and Chatterjee, E., Satyendra Nath Bose, (National Book Trust, India, 1976).

[7] Wali, K. C., Satyendra Nath Bose: His life and times, (World Scientific, Singapore, 2009).

[8] Anderson, M. H. J., Ensher, J. R., Metthews, M. R., Wieman, C. E., and Cornel, C. A., Observation of Bose-Einstein condensation in a dilute atomic vapour, Science, 1995, vol. **269**, p. 198-201.

[9] Bradley, C. C., Sackett, C. A., Tollett, J. J., and Hulet, R. G., Evidence of Bose-Einstein condensation in an atomic gas with attractive interaction, Phys. Rev. Lett., 1995, vol. **75**, p. 1687.

[10] Davis, K. B., Mewes, M. O., Andrews, M. R., van Druten, N. J., Durfee, D. S., Kurn, D. M. and Ketterle, W.,

Bose-Einstein condensation in a gas of sodium atoms, Phys. Rev. Lett., 1995, vol. **75**, p. 3969.

[11] Bose, S. N., Plancks Gesetz und Lichtquantenhypothese, Zeitschrift für Physik, 1924, vol. **26**, p. 178-181.

[12] Tyndall, J., On the Absorption and Radiation of Heat by Gases and Vapours, and on the Physical Connexion of Radiation, Absorption, and Conduction, Phil. Mag. and Journal of Science, 1861, vol. **22**, p. 169.

[13] Kirchhoff, G., On the relation between the radiating and absorbing powers of different bodies for light and heat, The London, Edinburgh, and Dublin Phil. Mag. and Journal of Science, 1860, vol. **20**, p. 1.

[14] Heilbron, J. L., A History of Atomic Models from the Discovery of the Electron to the Beginnings of Quantum Mechanics, diss. (University of California, Berkeley, 1964)

[15] Pathria, R. K., and Beale, P. D., Statistical Mechanics (3rd Ed., Elsevier, New York, 2021).

[16] Paschen, F., Uber gesetzmassigkeiten in den spectren fester korper, Annalen der Physik, 1896, vol. **294**, p. 455-492.

[17] Carson, T. R., Steps to the Planck function: A Centenary Reflection, arXiv: astro-ph/0011219v1 10 Nov 2000.

[18] Rayleigh, L., The dynamical theory of gases and radiation, Nature, 1905, vol. **72**, p. 54-55.

[19] Jeans, J. H., The dynamical theory of gases and radiation, Nature, 1905, vol. **72**, p. 101-102.

[20] Planck, M., On the law of the energy distribution in the normal spectrum, Ann. Phys., 1901, vol. **4**, p. 553 -563.

[21] Pais, A., Einstein and the quantum theory, Rev. Mod. Phys., 1979, vol. **51**, p. 863.

[22] Schiff, L. I., Quantum Mechanics, (Tata McGraw-Hill Edition, New Delhi, India, 2010).

[23] Saha, M. N. and Srivastava, B. N., A treatise on heat, (4th Ed., The Indian Press (Publication) Private Ltd., 1958).

[24] Einstein, A., Quantentheorie des einatomigen idealen Gases, Sitzungsberichte, 1924, p. 261-267.

[25] Sokol, P. E., Bose-Einstein condensation in Liquid Helium, edited by A. Griffin, D. W. Snoke and S. Stringari (Cambridge University press, Cambridge 1995) p. 51

[26] Silvera, I. F., Spin-Polarized Hydrogen: Prospects for Bose–Einstein Condensation and Two-Dimensional Superfluidity, edited by A. Griffin, D. W. Snoke and S. Stringari (Cambridge University press, Cambridge 1995) p. 160.

[27] Cohen-Tannoudji, C. N., Manipulating atoms with photons, Rev. Mod. Phys., 1998, vol. **70**, p. 707.

[28] Pethick, C. J., and Smith, H., Bose-Einstein Condensation in Dilute Gases (Cambridge University Press, Cambridge, 2004).

[29] Andrews, M. R., Mewes, M. -O., van Druten, N. J., Durfee, D. S., Kurn, D. R. and Ketterle, W., Direct, Nondestructive observation of a Bose-Condensate, Science, 1996, vol. **273**, p. 84 - 87.

[30] Andrews, M. R., Kurn, D. M., Miesner, H.-J., Durfee, D. S., Townsend, C. G., Inouye, S. and Ketterle, W., Propagation of sound in a Bose-Einstein Condensate. Phys. Rev. Lett. 1997, vol. **79**, p. 553.

[31] Hannaford, P. and Sacha, K., Condensed Matter Physics in Big Discrete Time Crystals, arXiv:2202.05544v1 [cond-mat.quant-gas] 11 Feb 2022.

[32] Fetter, A. L., Rotating trapped Bose-Einstein condensates, Rev. Mod. Phys., 2009, vol. **81**, p. 647.

[33] Dalibard J., Gerbier, F., Juzeliūnas, G. and Öhberg, P. Artificial gauge potentials for neutral atoms, Rev. Mod. Phys. 2011, vol. **83**, p. 1523.

[34] Bohm, A., Quantum Mechanics and Applications (3rd Edition, Springer-Verlag, 1993)

[35] Lin, Y.-J., Compton, R. L., Jiménez-García, K., Porto, J. V., and Spielman, I. B., Synthetic magnetic fields for ultracold neutral atoms, Nature, 2009, vol. **462**, p. 628-632.

[36] Lin, Y.-J. Compton, R. L., Jiménez-García, K., Phillips, W. D., Porto, J. V., and Spielman, I. B., A synthetic electric force acting on neutral atoms, Nature Physics, 2011, vol. **7**, p. 531-34.

[37] Y. -J. Lin, K. Jiménez-García and I. B. Spielman, Spin–orbit-coupled Bose–Einstein condensates, Nature, 2011, vol. **471**, p. 83-86.

[38] Bychkov, Yu. A. and Rashba, E. I., Oscillatory effects and the magnetic susceptibility of carriers in inversion layers, J. Phys. C: Solid State Phys., 1984, vol. **17**, p. 6039-6045.

[39] Dresselhaus, G., Spin-Orbit Coupling Effects in Zinc Blende Structures, Phys. Rev., 1955, vol. **100**, p. 580.

[40] M. Aidelsburger, M. Atala, S. Trotzky, Y. A. Chen and Bloch, I., Experimental realization of strong effective magnetic fields in an optical lattice, Phys. Rev. Lett., 2011, vol. **107**, p. 255301.

[41] Wang, P., Yu, Z. Q., Miao, Z. F., Huang, L., Chai, S., Zhai, H. and Zhang, J., Spin-orbit coupled degenerate Fermi gases, Phys. Rev. Lett., 2012, vol. **109**, p. 095301.

[42] Cheuk, L. W., Sommer, A. T., Hadzibabik, Z., Yefsah, T., Bakr, W. S. and Zwierlein, M. W., Spin-injection spectroscopy of a spin-orbit coupled Fermi gas, Phys. Rev. Lett., 2012, vol. **109**, p. 095302.

[43] Zhang, L. -Y., Ji, S. –C., Chen, Z., Zheng, J., Du, Z. -D., Yan, B., Pan, G. -S., Zhao, B., Deng, Y. -J., Zhai, H., Chen, S. and Pan, J. -W., Collective Dipole Oscillations of a spin-orbit coupled Bose-Einstein condensate, Phys. Rev. Lett. 2012, vol. **109**, p. 115301.

[44] Sinha, S., Nath, R. and Santos, L., Trapped two-dimensional condensates with synthetic spin-orbit coupling, Phys. Rev. Lett. 2011, vol. **107**, p. 270401.

[45] Ramachandhran, B., Opanchuk, B., Liu, X. –J. , Pu, H., Drummond, P. D., and Hu, H., Half-quantum vortex state in a spin-orbit-coupled Bose-Einstein condensate, Phys. Rev. A, 2012, vol. **85**, p. 023606.

[46] Li, Y., Martone, G. I. and Stringari, S., Sum rules, dipole oscillation and spin polarizability of a spin-orbit coupled quantum gas, Eur. Phys. Lett., 2012, vol. **99**, p. 56008.

[47] E. P. Gross, Hydrodynamics of a superfluid condensate, J. Math. Phys., 1963, no. 2, vol. **4**, p. 195-207.

[48] Pitaevskii, L. P., Vortex lines in an imperfect Bose gas, Sov. Phys. JETP, 196, vol. **13**, p. 451.

[49] Berge, L., Wave collapse in physics: Principles and applications to light and plasma waves. Phys. Rep.,1998, vol. **303**, p. 259.

[50] Vuong, L. T., Grow, T. D., Ishaaya, A.,. Gaeta, A. L, 't Hooft, G. W., Eliel, E. R., and Fibich, G., Collapse and l Vortices, Phys. Rev. Lett., 2006, vol. **96**, p. 133901.

[51] Y. V. Kartashov, L. Torner, M. Modugno, E. Ya Sherman, B. A. Malomed and V. V. Konotop, Multidimensional hybrid Bose-Einstein condensates stabilized by lower-dimensional spin-orbit coupling, Phys. Rev. Res., 2020, vol. **2**, p. 013036.

[52] Achilleos, V. , Frantzeskakis, D. J., Kevrekidis, P. G. and Palinovsky, D. E., Matter-Wave Bright Soliton in Spin-Orbit Coupled Bose-Einstein Condensates, Phys. Rev. Lett. 2013, vol. **110**, p. 264101.

[53] Xu, Y. , Zhang, Y. , and Wu, B., Bright solitons in spin-orbit coupled Bose-Einstein condensate, Phys. Rev. A, 2013, vol. **87**, p. 013614.

[54] R. P. Feynman, Statistical Mechanics, A Set of Lectures (Addison-Wesley Publishing Company, 1972).

[55] C. Hammer, Y. Zhang, M. A. Khamehchi, M. J. Davis and P. Engels, Spin-orbit coupled Bose- Einstein condensates in a one-dimensional optical lattice, Phys. Rev. Lett., 2015, vol. **114**, p. 070401.

[56] Eiermann, B., Anker, T., Albiez, M., Tagieber, M., Treulein, P., Marzlin K. P. , and Oberthalar, M. K., Bright Bose-Einstein gap solitons of atoms with repulsive interaction, Phys. Rev. Lett., 2004, vol. **92**, p. 230401.

[57] Anderson, B. P., and Kasevich, M. A., Macroscopic quantum interference from atomic tunnel arrays, Science, 1998, vol. **282**, p. 1686.

[58] Greiner, M., Mandal, O., Esslinger, T., Hansch, T. W., and Bloch, I., Quantum phase transition from a superfluid to a Mott insulator in a gas of ultracold atoms, Nature (London), 2002, vol. **415**, p. 39.

[59] Abdullaev, F. Kh., and Garnier, J., Propagation of matter-wave solitons in periodic and random nonlinear potentials, Phys. Rev. A, 2005, vol. **72**, p. 061605 (R).

[60] Morsch, O., and Oberthaler, M., Dynamics of Bose-Einstein condensates in optical lattices, Rev. Mod. Phys., 2006, vol. **78**, p. 176.

[61] Sekh, G. A. and Talukdar, B., Effects of optical lattices on bright solitons in spin-orbit coupled Bose-Einstein condensates, Phys. Lett. A, 2021, vol. **415**, p. 127665.

[62] N. Li, S. Elhatisari, E. Epelbaum, D. Lee, B. Lu and Ulf-G. Meißner, Galilean invariance restoration on the lattice, Phys. Rev. C, 2019, vol. **99**, p. 064001.

[63] P. Massella, F. Barranco, D. Lonardoni, A. Lovato, F. Pederiva and E. Vegezzi. Exact restoration of Galilei invariance in density functional calculations with quantum Monte Carlo, J. Phys. G. Nucl. Part. Phys., 2020, vol. **47**, p. 035105.

[64] Hassan, M. Z., and Kane, C. L., Colloquium: Topological insulators, Rev. Mod. Phys., 2010, vol. **82**, p. 3045.

[65] Qi, X. -L. and Zhang, S. -C. Topological insulators and superconductors, Rev. Mod. Phys., 2011, vol. **83**, p.1057.

[ 66] Wilczek, F., Majorana returns, Nat. Phys., 2009, vol. **5**, p. 614.

[67] von Klitzing, K., The quantized Hall effect , Rev. Mod. Phys., 1986, vol. **58**, p. 519.

[68] Flatte, E. M., IEEE Transaction on Electronic Devices, 2007, vol. **54**, no. 5, p. 907-920.

[69] Golam Ali Sekh, Dynamics of Matter-wave Solitons in Bose-Einstein Condendate, Ph. D Thesis (unpublished) Visva-Bharati University, Santiniketan, India (2010).

[70] Tin-Lun, Ho. and Zhang, S., Bose-Einstein condensation with spin-orbit interaction, Phys. Rev. Lett., 2011, vol. **107**, p. 150403.

[71] Romera, E. and Dehesa, J. S., The Fisher-Shannon information plane, an electron correlation tool, J. Chem. Phys., 2004, no. 9, vol. **120**, p. 8906-8912.

[72] Sekh, G.A., Saha, A. and Talukdar, B., Shannon entropies and Fisher information of K-shell electrons of neutral atoms, Phys. Lett. A, 2018, vol, **382**, p. 315-320.

[73] Sekh, G.A., Talukdar, B., Chatterjee, S. and Khan, B. A., Physica Scripta, 2022, vol. **97**, p. 115404.

[74] Shannon, C. E., A Mathematical Theory of Communication, Bell Syst. Tech. J, 1948, vol. **27**, p. 379.

[75] Fisher, R. A., Theory of Statistical Estimation, Proc. Cam. Phil. Soc., 1925, vol. **22**, p. 700.

[76] Bialynicki-Birula, I. and Myceilski, J., Uncertainty relations for information entropy in wave mechanics, Commun. Math. Phys., 1975, vol. **44**, p. 129.

[77] Hall, M. J. W., Quantum properties of classical Fisher information, Phys. Rev. A, 2000, vol. **62**, p. 012107.

[78] Dehesa, J. S., Gonzalez-Feres, R. and Sanchzes-Moreno, P., The Fisher information-based relation, Cramer-Rao inequality and kinetic energy for the D-dimensional central problem, J. Phys. A: Math. Theor., 2007, vol. **40**, p. 1845.

[79] Vadrel, V., Introduction to Quantum Information Science, (Oxford University Press, Oxford, 2006).

[80] Rabindranath Tagore (1861-1941), Indian poet; Nobel Prize in Literature, 1913, for *Gitanjali,* a collection of his lyric poems originally in Bengali, translated into English by Tagore himself and into French by Andre Gide.